# The Functional Wiener Filter

Benjamin Colburn, Luis G. Sanchez Giraldo, Jose C. Principe

**Abstract**

This paper presents a close form solution in Reproducing Kernel Hilbert Space (RKHS) for the famed Wiener filter, which we called the functional Wiener filter (FWF). Instead of using the Wiener-Hopf factorization theory, here we define a new lagged RKHS that embeds signal statistics based on the correntropy function. In essence, we extend Parzen's work on the autocorrelation function RKHS to nonlinear functional spaces. The FWF derivation is also quite different from kernel adaptive filtering (KAF) algorithms, which utilize a search approach. The analytic FWF solution is derived in the Gaussian kernel RKHS with a constant computational complexity similar to the Wiener solution, and never composes nor employs the error as in conventional optimal modeling. Because of the lack of congruence between the Gaussian RKHS and the space of time series, we compare performance of two pre-imaging algorithms: a fixed-point optimization (FWF$_{FP}$) that finds and approximate solution in the RKHS, and a local model implementation named FWF$_{LM}$. The experimental results show that the FWF performance is on par with the KAF for time series modeling, and it requires far less computation.

## Introduction

Norbert Wiener's 1949 work on minimum mean square error estimation opened the door for the theory of optimum filtering [1]. The mathematics to solve integral equations, the Wiener-Hopf method [2], were crucial to arrive at the optimal parameter function, however, the methodology is rather complex. In digital signal processing using finite impulse response filters, the Wiener solution coincides with least squares, as proven by the Wiener-Kinchin theorem [3]. Therefore, the solution still belongs to the span of the input data i.e., the corresponding filter is linear in the parameters and therefore it is not a universal functional approximator.

In the late 50's, Emmanuel Parzen [4] presented an alternative approach to solve the minimum mean square estimation (MMSE) problem in a Reproducing Kernel Hilbert space (RKHS) defined by the autocorrelation function of the data. Since RKHS theory will be extensively employed, we define here a RKHS. Let $E$ be a non-empty set, and $\kappa(u,v)$ a function defined on $E \times E$ that is nonnegative definite. Due to the Moore-Aronzsajn theorem [5], $\kappa(u,v)$ defines uniquely a RKHS, $\mathcal{H}_\kappa$, such that $\kappa(\cdot,v) \in \mathcal{H}_\kappa$ and for any $g \in \mathcal{H}_\kappa$, $\langle g, \kappa(\cdot,v) \rangle_{\mathcal{H}_\kappa} = g(v)$. Therefore, a RKHS is a special Hilbert vector space associated with a kernel such that it reproduces (via the inner product) in the space i.e., $\langle \kappa(\cdot,u), \kappa(\cdot,v) \rangle_{\mathcal{H}_\kappa} = \kappa(u,v)$; or equivalently, a space where every point evaluation functional is bounded. The history of RKHS applications started in physics [6], statistics [7], signal processing [8] and machine learning [12]. Here, it will also be clear that the RKHS framework provides a natural link between stochastic processes and deterministic functional analysis.

Parzen introduced for the first time the RKHS methodology in statistical signal-processing and time-series analysis in [4]. His essential idea is that there exists a congruence map between the set of random variables spanned by the random process $\{X(t), t \in T\}$ with covariance function $R(t,s) = E[X(t)X(s)]$ and the RKHS of vectors spanned by the set $\{R(\cdot,t), t \in T\}$ denoted as $\mathcal{H}_R$. Note that the kernel expresses the second-order statistics of the data through the expected value (a data-dependent kernel) and Parzen clearly stated that this RKHS offers an elegant functional analysis framework for minimum mean square error (MMSE) solutions such as regression coefficients, least squares estimation of random variables, detection of signals in Gaussian noise, and others [9],[10],[11]. Unfortunately, $\mathcal{H}_R$ is defined in the input data space, so yields only linear solutions to the regression problem. Parzen beautiful interpretation did not provide any practical improvement, so it was quickly forgotten in signal processing.

More recent work by Vapnik on support vector machines brought back a lot of interest to RKHS theory for pattern recognition [12], where the RKHS is used primarily as a high-dimensional feature space and the inner product is efficiently computed by means of the kernel trick. A nonnegative definite kernel function (e.g., Gaussian, Laplacian, polynomial, and others [13]) nonlinearly projects the data sample-by-sample into a high-dimensional RKHS. This development was included in adaptive filtering, yielding the class of kernel adaptive filters (KAF) [14], which allows the design of convex universal learning machines (CULMs) [15]. KAFs estimate a functional model that approximates the MMSE solution using search techniques in the RKHS defined by the Gaussian kernel [14], and the order grows linearly with the number of samples, if no sparsification is considered. Another branch of RKHS theory important for this paper is kernel Principal Component Analysis (KPCA) [16]. When the kernel function is infinite dimensional as the Gaussian, denoted as $G(x_i,.)$, the eigen decomposition of the empirical covariance operator $C = 1/N \sum_{i=1}^{N} G(x_i,.)G(x_i,.)^T$ needs to be truncated (we assume $G(x_i,.)$ are centered in the RKHS). In such cases, a more efficient approach uses only inner products of functionals centered at the projected samples, which can be computed in the input space using the reproducing property of the kernel (also called the kernel trick). The goal is to rewrite the eigen decomposition of the empirical covariance operator $C$ through a functional eigenvalue equation as $CV = \lambda V$, where $V$ is the eigenfunction $V = 1/N \sum_{i=1}^{N} \alpha_i G(x_i,.)$ and $\lambda$ is a vector of scalars that correspond to the eigenvalues. For any nonzero $\lambda$, the eigenfunction exists in the span of the RKHS defined by the kernel. Since the number of samples is finite this methodology is very appealing and efficient. However, the span of the functional space defined by the kernel is much larger than the mappings of single mapped samples into the RKHS, which means that the inverse mapping of RKHS functionals to the input space cannot be necessarily expressed as the image of a single input pattern i.e., given a function $\zeta$ in the RKHS span, there is no guarantee that there is exist a $z \in \mathbb{R}^N$ such that $G(z,.) = \zeta$. This has been called the preimage problem [17]. We call $\hat{z}$ an approximate preimage of $\zeta$ if $\|G(z,.) - \zeta\|^2$ is small, according to the application. We will see that this pre-imaging will be important in our approach.

This paper takes Parzen's work one step further, combining it with KAF concepts to yield a RKHS defined by the covariance function of the projected data in a Gaussian RKHS, which is nonlinearly related to the data space. More specifically, we define a data dependent kernel based on the correntropy function [16] that incorporates full data statistics and defines a RKHS of deterministic functions, even when the input data is a random variable (r.v.). Correntropy has been heavily used for robust cost functions in adaptive signal processing [17], but here its functional extension [16] will be employed as a methodology to solve the famous Wiener filter in the space of nonlinear functions, without using the Wiener-Hopf spectral factorization. Previous attempts by others e.g., the kernel Wiener filter [18], approximate the Wiener solution employing subspace projections in RKHS. An early attempt to solve the Wiener-Hopf equations in RKHS was not successful [19]. This paper shows how to pose the optimum filtering problem, derive a solution, and present a methodology to implement the filter directly from samples, which effectively extends MMSE for nonlinear universal approximators. The framework is named the functional Wiener filter (FWF) and amazingly, it does not require the use of the error signal as in the traditional

Wiener solution to adapt parameters. It takes advantage of the geometry of the RKHS and finds, just like Least Squares, the orthogonal projection of the desired response in the space spanned by the correntropy function, and in this sense, it is model agnostic. Preliminary results show that performance is on par with KLMS but it is worse than KRLS. The major advantage is the simplicity in implementation that is similar to the Wiener solution.

## Review of Linear Prediction of Continuous Time Series in RKHS

A stochastic process $X(t, \omega)$ is broadly defined as a collection of random variables on a measurable sample space $(\Omega, \mathcal{B}_\Omega)$, indexed by a set $T$. Here, we restrict $X(t, \omega)$ to random variables taking values in $\mathbb{R}$, $T \subset \mathbb{R}$, which we call a time-series, $\{X(t), t \in T\}$ and omit the dependence on $\omega$. For a time-series with finite second order moments, let $L_2(X(t), t \in T)$ denote the space of all real valued random variables spanned by the time series, that is, this space consists of all r.v. $U$ that are either linear combinations of finite number of $X(t_i)$ or are limits of such linear combinations. The time structure is quantified by the joint probability density $p_{t,s}(x_t, x_s)$ of the pair of random variables $X(t), X(s)$ at two points in time $t$ and $s$. Assuming a strictly stationary stochastic model for $X(t)$, the marginal density $p_t(x)$ is the same for any $t$. Normally, the joint density $p_{t,s}(x_t, x_s)$ is quantified by its mean value, called the autocorrelation function. To simplify notation, let us define the time autocorrelation of the finite second order moment time series as:

$$R(s, t) = E[X(s)X(t)] \quad (1)$$

This kernel on time sample pairs is positive semi definite, hence by Moore-Aronzsajn theorem [5] it defines a RKHS space of functions on $T \times T$, denoted $\mathcal{H}_R$. Notice that the functions in $\mathcal{H}_R$ are deterministic because of the $E[.]$ operator, while the inner product in $\mathcal{H}_R$ depends on the statistics of the data through $X(t)$.

For any r.v. $U, A \in L_2(X(t), t \in T)$, define the inner product between the two as

$$\langle U, A \rangle = E[UA], \quad (2)$$

and the norm of $U$ in $L_2(X(t), t \in T)$ by the inner product $\langle U, U \rangle = E[U^2]$. Obviously, this inner product coincides with the autocorrelation function if $U$ is $X(s)$ and $A$ is $X(t)$. However, notice that $L_2(X(t), t \in T)$ is not an RKHS.

Explicit Expression for MMSE
One of the important problems in time series analysis is the representation of an unobservable r.v. $Z$. Let $\{X(t), t \in T\}$ be an observable time series assumed stationary. The goal is to create a linear combination of the observable time series that has the smallest mean square distance to $Z$. By the Hilbert projection theorem, there is a unique minimum norm projection between the abstract Hilbert space $\mathcal{H}$ and any subspace $M$ of $\mathcal{H}$. Then, there exists a unique vector $A^*$ in $M$, given by $A^* = E^*[A|M]$, which projects orthogonally a vector $A$ in $\mathcal{H}$ to $M$. For a family of vectors $\{X(t), t \in T\}$ the projection becomes

$$A^* = E^*[A|X(t), t \in T]. \quad (3)$$

Then with $A = Z$, the optimum linear predictor is the unique r.v. in $L_2(X(t), t \in T)$ that satisfies

$$E[E^*[Z|X(t), t \in T]X(s)] = E[ZX(s)] \quad (4)$$

This result gives immediately rise to the famous Wiener equation. Indeed, if $T$ is a finite interval $T = \{t: a \leq t \leq b\}$ and $w(t)$ a weighting function in $L_2$, the integral

$$\int_a^b X(t)w(t)dt \quad (5)$$

represents a r.v. in $L_2(X(t), t \in T)$, then the weighting function of the best linear predictor can be written as

$$E^*[Z|X(t), t \in T] = \int_a^b X(t)w^*(t)dt \quad (6)$$

and must satisfy the generalized Wiener equation:

$$\int_a^b w^*(t)R(s,t)dt = \rho_z(s) \quad (7)$$

in $a \leq s \leq b$, with $R(s,t) = E[X(s), X(t)]$, $\rho_z(s) = E[ZX(s)]$.

This equation states that one can always find a representation for the function $\rho_z(s)$ in terms of the functions $\{R(s,t), t \in T\}$ such that the minimum mean square error linear predictor $E^*[Z|X(t), t \in T]$ can be written in terms of the corresponding linear operator on the time series $\{X(t), t \in T\}$.

Hilbert Space Representation of Time Series
First, let us state an important theorem that is very important for this line of work [4].

*Basic Congruence Theorem.* Let $\mathcal{H}_1$ and $\mathcal{H}_2$ be two abstract Hilbert spaces. Let $T$ be an index set and let $\{u_t, t \in T\}$ be a family of vectors spanning $\mathcal{H}_1$, and similarly $\{a_t, t \in T\}$ a family of vectors spanning $\mathcal{H}_2$. Suppose that for every $s$, $t$ in $T$,

$$\langle u_s, u_t \rangle_{\mathcal{H}_1} = \langle a_s, a_t \rangle_{\mathcal{H}_2} \quad (8)$$

then there is a congruence (a one-to-one inner product preserving linear mapping) $\psi$ from $\mathcal{H}_1$ to $\mathcal{H}_2$ such that $\psi(u_t) = a_t$ for any $t \in T$.

*Definition*: A family of vectors $\{u_t, t \in T\}$ in a Hilbert space $\mathcal{H}_R$ is a representation of a wide sense stationary time series $\{X(t), t \in T\}$ if for every $s$, $t$ in $T$

$$\langle u_s, u_t \rangle_{\mathcal{H}_R} = R(s,t) = E[X(s), X(t)] \quad (9)$$

Then there is a congruence $\psi$ between the Hilbert space spanned by $\{u_t, t \in T\}$ and denoted as $L_2(u_t, t \in T)$, onto $L_2(X(t), t \in T)$ satisfying $\psi(u_t) = X(t)$, and every r.v. $U$ in $L_2(X(t), t \in T)$ may be written $U = \psi(g)$ for some unique vector $g$ in $L_2(u_t, t \in T)$.

The natural representation of a time series is obtained in the RKHS $\mathcal{H}_R$ i.e., a Hilbert space where the kernel has two properties:

$$R(\cdot, t) \in \mathcal{H}_R$$
$$\langle g, R(\cdot, t) \rangle_{\mathcal{H}_R} = g(t) \quad (10)$$

This result is the well-known Riez representation theorem, which yields for our discussion

$$R(s, t) = \langle R(\cdot, s), R(\cdot, t) \rangle_{\mathcal{H}_R} = E[X(s), X(t)] \quad (11)$$

It can be further shown that for any time series $\{X(t), t \in T\}$ with covariance kernel $R$, the family of functions $\{R(\cdot, t), t \in T\}$ in $\mathcal{H}_R$ is a representation of $L_2(X(t), t \in T)$. Indeed, for any two vectors $U, A \in L_2(X(t), t \in T)$ such that the congruence is denoted by $U = \psi(g)$ and $A = \psi(h)$, and $A = \psi(h) = \langle X, h \rangle_{\mathcal{H}_R}$,

$$\langle X, R(\cdot, t) \rangle_{\mathcal{H}_R} = X(t)$$
$$E[\langle X, h \rangle_{\mathcal{H}_R} \langle X, g \rangle_{\mathcal{H}_R}] = \langle h, g \rangle_{\mathcal{H}_R}$$

It is easy to see that if the two vectors $h, g \in \mathcal{H}_R$ correspond to random variables $U, A \in L_2(X(t), t \in T)$

$$\langle h, g \rangle_{\mathcal{H}_R} = \int_{t \in T} \int_{s \in T} h(s) R^{-1}(s, t) g(t) \, ds \, dt,$$

where $R^{-1}(s, t)$ is the kernel of the inverse of the covariance operator $Rg = \int_{t \in T} g(t) R(s, t) dt$. Moreover, if $U = \sum_{i=1}^{N_g} w_g(t_i) X(t_i)$ and $A = \sum_{j=1}^{N_h} w_h(s_j) X(s_j)$, their inner product in the RKHS can be computed in the input space from vectors $\{w_h(s_j)\}_{j=1}^{N_h}$ and $\{w_g(t_i)\}_{i=1}^{N_g}$ (what is now called the kernel trick) as

$$\langle h, g \rangle_{\mathcal{H}_R} = \sum_{j=1}^{N_h} \sum_{i=1}^{N_g} w_h(s_j) R(s_j, t_i) w_g(t_i) = \sum_{j=1}^{N_h} \sum_{i=1}^{N_g} h(s_j) r_{s_j, t_i}^{-1} g(t_i) \quad (13)$$

where $r_{s_j, t_i}^{-1}$ is the $s_j, t_i$ element of the inverse of the covariance kernel $R(s_i, t_i)$ i.e., the kernel modifies the traditional inner product of vectors in the input space. This explains the nature of $\mathcal{H}_R$ quite well: because of the mapping $R(s, .)$, which contains the statistics of the data, the inner product in $\mathcal{H}_R$ takes advantage of the data statistics over time instances. Hence, in the input space the solution must be a quadratic form employing $R^{-1}$ as shown in (13) to meet the congruence.

*Theorem (from [4])*: Let $\{X(t), t \in T\}$ be a time series with covariance kernel $R(s, t)$, and let $\mathcal{H}_R$ be the corresponding RKHS. Between $L_2(X(t), t \in T)$ and $\mathcal{H}_R$ there exists a one-to-one inner product preserving linear mapping under which a vector $h \in \{R(\cdot, t), t \in T\}$ and $U \in L_2(X(t), t \in T)$ are mapped into one another. Denote by $\langle h, X \rangle_{\mathcal{H}_R}$ the r.v. in $L_2(x_t, t \in T)$ which corresponds to the function $h \in \mathcal{H}_R$ under the mapping. Then the solution of the prediction problem may be written as follows. If $Z$ is a r.v. with finite second moments and $\rho_Z(t) = E[ZX(t)]$ then
$$\rho_Z \in \mathcal{H}_R, \text{ and} \quad (14)$$

$$E^*[Z|X(t), t \in T] = \langle \rho_z, X \rangle_{\mathcal{H}_R} \qquad (15)$$

with prediction mean square error given by

$$E[(Z - E^*[Z|X(t), t \in T])^2] = E[Z^2] - \langle \rho_z, \rho_z \rangle_{\mathcal{H}_R} \qquad (16)$$

The equivalent minimum mean square error solution (15) in the data space, because of (13), becomes

$$Y = E^*[Z|X(t), t \in T] = \langle \rho_z, X \rangle_{\mathcal{H}_R} = \int_{t \in T} \int_{s \in T} R^{-1}(s,t) \rho_Z(s) X(t) ds\, dt \qquad (17)$$

which is exactly the Wiener solution $\int_{t \in T} X(t) w^*(t) dt$. Note that the effective role of this inverse operator is to decorrelate the input space data and it is a steppingstone for finding the orthogonal projection as demonstrated by Wiener. However, in $\mathcal{H}_R$ this solution for the prediction problem is coordinate free, *does not use the approximation error*, and directly uses the structure of $\mathcal{H}_R$. In fact, it is sufficient to compute the linear projection of $\rho_z(s)$ with the input data because the covariance kernel $R(s,t)$ provides its statistics, unlike Wiener-Hopf method that requires spectral factorization. This coordinate free property of RKHS solutions with the covariance kernel was first noted by Loeve [20] who suggested that instead of finding a set of functional projections (e.g. Karuhnen Loeve transform [21]) it is sufficient to employ the statistics of $X(t)$ embedded in the structure of the RKHS. Parzen [4] further states that for this reason "*RKHS defined by the covariance kernel is the natural setting in which to solve problems of statistical inference on time series*". These are fundamental results that will be very useful when seeking an extension of the theory to nonlinear solutions.

The fundamental issue with Parzen approach is twofold: first, it does not elucidate efficient alternatives to implement the conditional mean operator. Moreover, from (13) we can see that the inverse may not always exist, needs to be accurate, and it is computationally expensive because it needs to be applied to every test sample. Despite approximations for the inverse, this is cumbersome but a necessity for continuous time models. Second, for discrete time signal processing, this approach is computationally not competitive with the famous Wiener solution $w^* = R^{-1}\rho$, where $R$ is the autocorrelation matrix (the kernel $R(s,t)$ evaluated at a finite set of times), which finds the optimal weighting $w^*$ only once in the training set using the error, and does an inner product in the data space of two vectors in the test set. Hence, we conclude that the advantage of the RKHS theory is on the mathematical tools of congruence and representation of time series in RKHS, which open the door to seek more general solutions such as the nonlinear prediction case. In fact, the advantage of the RKHS theory is that the operations defined in the RKHS are independent of the kernel utilized, hence the key goal is to concentrate on designing proper kernels when the goal is nonlinear extensions.

**The Nonlinear Prediction Case**
    **A. Kernel Adaptive Filtering**

The goal is to construct a function $f: \mathbb{R}^L \to \mathbb{R}$ based on a real sequence $\{(x_i, d_i)\}_{i=1}^N$ of examples $(x_i, d_i) \in S \times D$, where $D$ is a compact subset of $\mathbb{R}$ and $S$ a compact subspace of $\mathbb{R}^L$. As described below, the function $f \in \mathcal{H}_k$ is obtained based on a positive definite kernel $\kappa: S \times S \to \mathbb{R}$ that defines a RKHS $\mathcal{H}_k$. A commonly employed kernel is the Gaussian kernel $G(x, x_i) =$

$\exp\left(-\frac{\|x-x_i\|^2}{2\sigma^2}\right)$, where $\sigma$ is the kernel size or bandwidth. Kernel adaptive filtering (KAF) [14] implements nonlinear filtering on discrete time series by mapping the input sampled data $\{x_i\}_{i=1}^N$ to $\mathcal{H}_G$ using a positive definite kernel $G$, and using search techniques based on the gradient and or Hessian information to adapt functional parameters.

The Gaussian kernel maps each embedding vector $x_i$ of size $L$, to a function in $\mathcal{H}_G$, which we will also denote as $G(x_i,\cdot)$, where the "·" in the second argument means that a data point is represented by a Gaussian function centered at $x_i$. The inner product in the RKHS of two such functions centered at $x_i$ and $x_j$ can be easily computed in the input space as a Gaussian kernel evaluation i.e., $\langle G(x_i,\cdot), G(x_j,\cdot)\rangle_{\mathcal{H}_G} = G(x_i, x_j)$. The $\mathcal{H}_G$ defined by the Gaussian is infinite dimensional and nonlinearly related to the input data space $S$ [22]. For the case of samples from a stochastic process $\{X(t), t \in T\}$, $G(X(t),\cdot)$ is a random function. One notable example of KAF is the kernel least mean square (KLMS) algorithm, for which the non-linear filter output is simply given by

$$y_n = \sum_{i=1}^{n-1} \eta e_i G(x_n - x_i) \quad (18)$$

where $\eta$ is the stepsize, $e_i$ is the error at iteration $i$, and $\{x_i\}_{i=1}^{n-1}$ are the past samples in the training set that constitute the "dictionary" to construct the output. This algorithm uses gradient search to construct the optimal function $\Omega^*$, such that $f^*(x) = \langle G(x,\cdot), \Omega^*\rangle_{\mathcal{H}_G}$, and converges in the mean to the optimal least minimum square solution in $\mathcal{H}_G$ for small step sizes and large number of data samples. The appeal of the KLMS is that it is an online algorithm, does not need explicit regularization [23], and is a CULM (convex and universal learning machine) [15]. However, because of the nonlinearity of the kernel mapping there is no congruence between the input space defined by the span of the time series and the RKHS $\mathcal{H}_G$. The solution needs to be expressed in terms of observations from the time series, which means that the order of the filter grows linearly in time, if no sparsification is included [14]. This is a shortcoming of this class of algorithms because it affects the computation complexity in the test set. In KAF, since the kernel evaluations are weighted by the error, the algorithm has an automatic way to preserve the scale of the representations when applying the kernel trick.

The $\mathcal{H}_G$ defined by the Gaussian kernel differs from the $\mathcal{H}_R$ defined by Parzen's covariance kernel in four fundamental ways.

- First, Parzen used a "linear" kernel $\mathcal{H}_R$ yielding a close form optimal linear model in $L_2$ as mentioned above.
- Second, the Parzen kernel is computed by employing the expected value over data lags $s = t - \tau$ to take advantage of the wide sense stationarity of the time series, unlike the pairwise sample set as $\mathcal{H}_G$.
- Third, the map to $\mathcal{H}_G$ is stochastic because samples are mapped from a random process rather than mapping the elements of the index set $T$, directly. In contrast, the map to $\mathcal{H}_R$ is deterministic because of the congruence.
- Fourth, $\mathcal{H}_G$ is infinite dimensional, while in $\mathcal{H}_R$ is a finite dimensional RKHS space defined by the number of lags required for the covariance kernel, which is dictated by the input data dynamics (normally small).

Our goal now is to define a new RKHS that preserves the correlation structure defined by the data as $\mathcal{H}_R$, but also maps the time series by a nonlinear kernel to achieve CULM properties. To be practical, this approach uses the kernel trick to perform the computation in the input space.

## B. Definition of the Correntropy RKHS

Let $\{X(t), t \in T\}$ be a strictly stationary stochastic process (i.e., the joint PDF $\{p_{s,t}(x_s, x_t)\}$ is unaffected by a change of the time origin, that is $p_{s,t}(x_s, x_t) = p_{s-\tau,t-\tau}(x_s, x_t)$) with $T$ being an index set and $x_t \in \mathbb{R}^L$. The autocorrentropy function $v(s,t)$ is defined as a function from $T \times T \to \mathbb{R}$ given by

$$v_\sigma(s,t) = E_{s,t}[G_\sigma(X(s), X(t))] = \iint G_\sigma(x_s, x_t) p_{s,t}(x_s, x_t) dx_s dx_t \quad (19)$$

where $E_{s,t}[\cdot]$ denotes mathematical expectation over a pair of r.v. in the time series $\{X(t), t \in T\}$. While it is true that any symmetric positive definite kernel (i.e., Mercer kernel) $\kappa(x_s, x_t)$ can be employed instead of the Gaussian kernel $G_\sigma$, the symmetry, scaling, and translation invariant properties of $G_\sigma$, confer additional properties and interpretation to correntropy, which are reviewed in the appendix. The autocorrentropy function defined in (19) is a reproducing kernel on the index set $T \times T$. We will denote its corresponding RKHS by $\mathcal{H}_v$. The functions $v_\sigma(s, \cdot)$ are in $\mathcal{H}_v$ and $v_\sigma(s,t) = \langle v_\sigma(s, \cdot), v_\sigma(t, \cdot) \rangle_{\mathcal{H}_v}$.

Another space that can be defined by the composition of the random variable $X(t)$ and the positive definite Gaussian kernel $G_\sigma(\cdot, \cdot)$ is the span of the set of random elements $\{G_\sigma(X(t), \cdot), t \in T\}$ taking values in $\mathcal{H}_G$. We will denote this space by $\mathcal{H}_{RG}$ and the inner product between two elements $U = \sum_i \alpha_i G_\sigma(X(t_i), \cdot)$ and $A = \sum_j \beta_j G_\sigma(X(s_j), \cdot)$ is given by

$$\langle U, A \rangle_{\mathcal{H}_{RG}} = E\big[\langle \sum_i \alpha_i G_\sigma(X(t_i), \cdot), \sum_j \beta_j G_\sigma(X(s_j), \cdot) \rangle_{\mathcal{H}_{RG}}\big] = \sum_{ij} \alpha_i \beta_j E\big[G_\sigma(X(t_i), X(s_j))\big].$$

There is a congruence between $\mathcal{H}_{RG}$ and $\mathcal{H}_v$. Moreover, we see that for strictly stationary time series making $s = t - \tau$, the function $v_\sigma$ can also be written as a function of $\tau$ only as follows:

$$v_\sigma(\tau) = E_{t,t-\tau}[G_\sigma(X(t), X(t-\tau))], \quad (20)$$

where any $t \in T$ can be used. This shows its similarity with the Parzen covariance kernel of (11), except that $v_\sigma(\tau)$ is computed in $\mathcal{H}_v$, a space nonlinearly related to the original time series.

The autocorrentropy functional can then be interpreted in two vastly different feature spaces. One is the RKHS $\mathcal{H}_G$ induced by the Gaussian kernel on pairs of observations $G_\sigma(\cdot, \cdot)$, which is widely used in kernel learning. The elements of this RKHS are infinite-dimensional vectors expressed by the eigenfunctions of the Gaussian kernel, and they lie on the positive hyperoctant of a sphere because $\|G_\sigma(x, .)\|^2 = G_\sigma(0) = 1/\sqrt{2\pi}\sigma$. The correntropy functional performs statistical averages on the functionals in this sphere.

The second feature space is the RKHS $\mathcal{H}_v$ induced by the correntropy kernel $v(s,t)$, which is defined on the index set of the random variables in the time series. This inner product is defined by the correlation of the kernel at two different lags and the mapping produces a single deterministic scalar for each element on the index set, that is, the practical dimension of $\mathcal{H}_v$ is the size of the index set. $\mathcal{H}_v$ has very nice properties for statistical signal processing:

- $\mathcal{H}_v$ provides a straightforward way to apply optimal projection algorithms based on mean statistical embeddings that are expressed by inner products.
- The effective dimension of $\mathcal{H}_v$ is under the control of the designer by selecting the number

of lags (just like with the RKHS defined by the autocorrelation function).
- Elements of $\mathcal{H}_v$ can be readily manipulated algebraically for statistical inference (i.e. without taking averages over realizations).
- $\mathcal{H}_v$ is nonlinearly related to the input space, unlike the RKHS defined by the autocorrelation of the random process. Therefore, it is in principle very appealing for nonlinear statistical signal processing.

The table presents the different types of RKHS defined so far that summarize our approach.

Table I

| RKHS | Functional Mapping | Hilbert Space Characteristics |
|---|---|---|
| $\mathcal{H}_R$ Parzen | $E[X(t),.]$ | Linear mapping of data, size of lags, deterministic functions |
| $\mathcal{H}_G$ Gaussian | $G(x,\cdot)$ | Nonlinear mappings of data, infinite dimensional, random functions |
| $\mathcal{H}_{RG}$ Random Gaussian | $G(X(t),\cdot)$ | Nonlinear mapping of data, size of lags, random functions |
| $\mathcal{H}_v$ Correntropy | $v_\sigma(t,\cdot)$ | Nonlinear mapping of data, size of lags, deterministic functions |

Representing an Unobservable Random Variable in $\mathcal{H}_{RG}$

Like the original problem where the random variable $Z$ was approximated by a random variable in the span of the time series $\{X(t), t \in T\}$ by the Hilbert projection theorem, we can define the approximation in the space of random elements $\mathcal{H}_G$ as follows

$$\xi^* = \underset{\xi}{\operatorname{argmin}} E\big[\|G(Z,\cdot) - \xi\|_{\mathcal{H}_G}^2\big], \qquad (21)$$

where $\xi$ is a random element in the span of $\{G(X(t),\cdot), t \in T\}$. Solving for $\xi$ gives rise to the following equation:

$$E\big[\langle G_\sigma(Z,\cdot), G_\sigma(X(s),\cdot)\rangle_{\mathcal{H}_G}\big] = E\big[\langle \xi, G_\sigma(X(s),\cdot)\rangle_{\mathcal{H}_G}\big], \qquad (22)$$

where $\xi$ is expressed as a linear combination of elements in $\mathcal{H}_{RG}$,

$$\xi = \int_T G_\sigma(X(t),\cdot)w(t)dt, \qquad (23)$$

Then the weighting function $w^*$ of the best predictor must satisfy:

$$E\Big[\int_T w^*(t)\langle G_\sigma(X(t),\cdot), G_\sigma(X(s),\cdot)\rangle_{\mathcal{H}_G} dt\Big] = E\big[\langle G_\sigma(Z,\cdot), G_\sigma(X(s),\cdot)\rangle_{\mathcal{H}_G}\big],$$

which gives rise to the functional Wiener equation:

$$\int_T w^*(t) v_\sigma(t,s) dt = E\big[\langle G_\sigma(Z,\cdot), G_\sigma(X(s),\cdot)\rangle_{\mathcal{H}_G}\big] = \rho_Z(s), \qquad (24)$$

These equations state that one can always find a representation for the function $\rho_z(s)$ in terms of the functions $\{v_\sigma(t,\cdot), t \in T\}$ because the best correntropy predictor is computed in the span of the set $\{G_\sigma(X(t),\cdot), t \in T\}$. Nevertheless, because this computation is carried out in the correntropy RKHS, the best approximation to $Z$ cannot be directly obtained since the input space where the time series lies is nonlinearly related to the correntropy RKHS where we compute the projection.

Solution of the Representation Problem in $\mathcal{H}_v$

To solve the representation problem in $\mathcal{H}_v$, that is, finding $w^*(t)$, let us consider the representation $\zeta_s$ in $\mathcal{H}_v$ of the random element $G_\sigma(X(s),\cdot)$ that can be obtained by the congruence between $\mathcal{H}_v$ and $\mathcal{H}_G$. From equation (24) we have that:

$$\rho_z(s) = \langle \rho_z, \zeta_s \rangle_{\mathcal{H}_v} = \int_T w^*(t) \langle \zeta_t, \zeta_s \rangle_{\mathcal{H}_v} dt \qquad (25)$$

This defines a close form functional Wiener filter solution in $\mathcal{H}_v$. Notice that the formulation is the same as (17), the only difference is the structure of the inner product space.

The relation between $\mathcal{H}_{RG}$ and $\mathcal{H}_v$ is rather similar to the relation between $\mathbb{R}^L$ and $\mathcal{H}_R$ so, for two elements $h$ and $g$ in $\mathcal{H}_v$,

$$\langle h, g \rangle_{\mathcal{H}_v} = \int_{t \in T} \int_{s \in T} h(s) v_\sigma^{-1}(s,t) g(t) ds\, dt, \qquad (26)$$

where $v_\sigma^{-1}(s,t)$ is the element of the inverse of the correntropy operator defined as $(V_\sigma g)(s) = \int_{t \in T} g(t) v_\sigma(s,t) dt$. The above form can be used to compute a solution to (25) as,

$$w^*(t) = \int_{s \in T} \rho_z(s) v_\sigma^{-1}(s,t) ds. \qquad (27)$$

In this case the solution is nonlinear in the input space, so this is a very elegant extension of Wiener theory. A major difference to KAF and the Wiener filter in the data space, is that this solution never uses the error. The reason is that Parzen's solution decorrelates implicitly the data (in this case in $\mathcal{H}_{RG}$) and automatically finds the orthogonal projection on the data manifold.

However, not everything is perfect with the solution (26), since we cannot extend the congruence in (25) to the original time series $\{X(t), t \in T\}$, i.e.

$$\langle \zeta_t, \zeta_s \rangle_{\mathcal{H}_v} = E[G_\sigma(X(t), X(s))] \neq E[X(t)X(s)] \qquad (28)$$

because the kernel mapping does not preserve the inner product, i.e. $\langle x_n, x_i \rangle \neq \langle G(x_n,.), G(x_i,.) \rangle_{\mathcal{H}_G}$.

### C. Computation of the Functional Wiener Filter in $\mathcal{H}_G$

How can the solution in (26) be implemented from a sample data stream? In this case, we restrict our treatment to discrete-time time series. Let us start by assuming that the time series is ergodic, such that expected values can be estimated by temporal averages. Second, because of the congruence (25), $\langle \zeta_t, \zeta_{t-\tau} \rangle_{\mathcal{H}_v}$ can be substituted by $E[G_\sigma(X(t), X(t-\tau))]$ and by ergodicity, it can be estimated from samples $\{x(t)\}_{t=1}^N$ over a window of length $N$.

$$v_\tau = \frac{1}{N}\sum_{t=1}^{N} G_\sigma(x(t), x(t-\tau)) \qquad (29)$$

For $\tau = 0, 1, \cdots, L-1$, $v_\tau$ is the $\tau^{\text{th}}$ entry of the autocorrentropy vector and can be used to construct the autocorrentropy matrix of size $L \times L$ as follows:

$$V = \begin{bmatrix} v_0 & \cdots & v_{T-1} \\ \vdots & \ddots & \vdots \\ v_{T-1} & \cdots & v_0 \end{bmatrix} \qquad (30)$$

This matrix is unlike anything in kernel adaptive filtering, because it is a matrix of scalar values that can be computed once from the training set and never changed. This matrix is very unusual in kernel filtering, where the filters always increase in size with each new sample. The values of the correntropy matrix can be centered in RKHS if necessary [28]:

$$\bar{v}_\tau = v_\tau - \frac{1}{N^2}\sum_{t=1}^{N}\sum_{s=1}^{N} G_\sigma(x(t), x(s)) \qquad (31)$$

The other major difference is that in KAF, one needs to transfer vectors of samples to the RKHS, where the size of the vector is an estimate of the embedding dimension of the system that created the time series, using Takens' embedding theory. The reason is that the KLMS is a pairwise instantaneous algorithm, so if it is applied to each sample of the input data the algorithm loses the local time structure of the signal. For FWF, the data can be mapped to RKHS sample by sample, just like in the input space, because the formulation uses the correntropy matrix where the lag structure is included.

Let us now show how to estimate the cross correlation functional $\rho_z$ in $\mathcal{H}_v$. Using the same approximations as the ones for the correntropy matrix yields

$$\hat{\rho}_z(\tau) = \frac{1}{N}\sum_{t=1}^{N} G_\sigma(x(t-\tau), z(t)) \qquad (32)$$

This is the only term that relates the target and the input signals, and it only needs to be evaluated in the training set. The optimal weighting vector in (27), $w^*(\tau)$ for $\tau = 0, 2, \ldots, L-1$, is obtained by solving the system:

$$\rho_z(\ell) = \sum_{\tau=0}^{L-1} V_{\ell+1,\tau+1}\, w(\tau). \qquad (33)$$

In other words, $w^* = V^{-1}\rho_z$. During testing, the output of the filter corresponds to an instance of the random element $\sum_{\tau=0}^{L-1} w^*(\tau) G_\sigma(X(t-\tau), \cdot)$, which is the best approximation to $G_\sigma(Z, \cdot)$, namely,

$$\xi^*(t) = \sum_{\tau=0}^{L-1} w^*(\tau) G_\sigma(x_{\text{test}}(t-\tau), \cdot). \qquad (34)$$

where $x_{\text{test}}(t)$ is the test input at time $t$. This solution shares the form of (6) in $L_2(X(t), t \in T)$ and (23) in $\mathcal{H}_{RG}$. The big difference is that the autocorrelation function was substituted by the correntropy function, while the input vector $[x(t), x(t-1), \cdots, x(t-L+1)]$ was substituted by a vector of functions nonlinearly related to the input space (the feature space defined by the Gaussian kernel).

Notice that this solution is quite different from the KAF in several important ways. First, the optimal weight vector can be computed in the input space, and it appears as a scale factor to change the finite range of the Gaussian to span the values of the target response. Notice that this weighting depends on the actual local $L$ sample history of the current input, but it is nonlinear and so it is more powerful than the linear weighting in linear Wiener filters. Second, there is no sum over the training set samples in the optimal solution like in KAFs. The best approximant is a combination of just $L$ Gaussian functions centered at the current test sample, which is a major simplification in computation when compared with KAF. This algorithm has the complexity of the Wiener solution, and should be an universal approximator when the number of delays grows to infinity, but we have not formally proved this statement. Unfortunately, the output of the functional Wiener filter $\xi^*(t)$ is still in $\mathcal{H}_G$, so the task of implementing a filter in the data space is still not finalized.

### D. Preimage to Estimate the FWF output in the input space

Ideally, the output of the FWF in the input space would correspond to the inverse map from $\mathcal{H}_G$ to $\mathbb{R}^d$, where $d = 1$ in the simplest. Since (34) expresses the optimal filter solution as a linear combination of Gaussian function, the goal is just to evaluate the function at a point in the input space, whose image is closest to the optimal solution. However, there is no guarantee such inverse map exists, so we must resort to an extra optimization or approximation step to find a pre-image [17] of the optimal solution in the input space, as will be explained next.

#### D.1. Preimage using the optimal filter output in $\mathcal{H}_G$

For the FWF, the basic concept is to use an approximate pre-image in the input space of the optimal filter output in $\mathcal{H}_G$ i.e., the approximated FWF output to $y^*(t)$ will be given by:

$$y(t) = \underset{y \in \mathbb{R}^d}{\operatorname{argmin}} \|G_\sigma(y,\cdot) - \xi^*(t)\|^2_{\mathcal{H}_G} \qquad (35)$$

This formulation can be applied in practical settings because in a training set, the optimal weight vector can be estimated using the $V$ matrix from (33) and the cross correntropy from (32). Therefore, and according to (35) it is only required to find the point to evaluate the optimal weight function, which is equivalent to find the minimum of

$$\sum_{\tau=0}^{L-1} w^*(\tau) G_\sigma(x_{\text{test}}(t-\tau), y). \qquad (36)$$

Making the gradient of (36) with respect to $y$ equal to zero yields the fixed-point expression

$$y^{(i+1)} = \frac{\sum_{\tau=0}^{L-1} w^*(\tau) G_\sigma(x_{\text{test}}(t-\tau), y^{(i)}) x_{\text{test}}(t-\tau)}{\sum_{\tau=0}^{L-1} w^*(\tau) G_\sigma(x_{\text{test}}(t-\tau), y^{(i)})}, \qquad (37)$$

where $y^{(i)}$ denotes the estimate of the preimage at the $i$th iteration of the fixed-point update. Notice that the nature of the pre-imaging solution involves a search on top of the analytic solution. This solution will be named FWF$_{FP}$.

### D.2. Preimage using local models

Intuitively, the goal is to select training set input samples that, when combined with the current test sample, provide functional evaluations in RKHS that approximate the targets in the training set. The difficulty is that during testing there is no information about the target value. Therefore, one simple option is to use the similarity in the input space to cluster locally the input samples that provide the best approximation to the target signal during training. This approach was inspired by [29], where a successful table lookup approach was employed to extend linear model performance that links input samples to their errors in the training set to create outputs outside the span of the input space.

Here, the approach is to find an input sample $x(m)$ that when combined with the current input $x(i)$, will produce an output in $\mathcal{H}_G$ that is close to its target $z(i)$. Let us represent $\hat{z}(i) = \sum_{\tau=0}^{L-1} w^*(\tau) G_\sigma(x(i-\tau), x(m-\tau))$. The optimization can be written as

$$arg \min_{x(m) \in S} \|z(i) - \hat{z}(i)\| \quad (38)$$

where $S$ is the training set. So, we need to implement a search (done once), where we find the sample pair $(x(i), x(m)), i = 1, \dots N$ that produces the closest approximation to the target sample $z(i)$. Once in testing, we find the closest sample $x(i)$ in the training set to $x(test)$ and use its neighbor $x(m)$ to plug in (34) to obtain the FWF output as

$$y(t) = \frac{z_i}{\hat{z}_i} \sum_{\tau=0}^{L-1} w^*(\tau) G_\sigma(x(m-\tau), x(t)) \quad (39)$$

where the ratio $z_i/\hat{z}_i$ enforces the scale. This search needs to be done online for every test sample, but if we rank the training set, it can be done quickly with a tree search. This process can be repeated K times for a better approximation, where K is a hyper-parameter. The idea is to probe the neighborhood of $x(test)$ with K input samples $\{x(1), \dots x(K)\}$ and use their respective neighbors using (38) to compute K approximate targets $\{\hat{z}(1), \dots \hat{z}(K)\}$ and represent their mean by $\bar{z}$. The final FWF output will be

$$y(t) = \sum_{k=1}^{K} \frac{z_k}{\bar{z}} \sum_{\tau=0}^{L-1} w^*(\tau) G_\sigma(x(k-\tau), x(t)) \quad (40)$$

Since the filter computation is so small, this improves performance with a minor increase in computation. The computational complexity of FWF$_{FP}$ and FWF$_{LM}$ are compared in the following table (i = iterations, M = fixed point updates).

Table II

| Filter | Complexity (training/testing) | Memory (training/ testing) |
| --- | --- | --- |
| KLMS | O(i) | O(i) |
| KRLS | O(i$^2$) | O(i$^2$) |
| FWF$_{FP}$ | O(L$^2$N)/ O(L) + O(LM) | O(N+L$^2$)/O(L) |
| FWF$_{LM}$ | O(L$^2$N) + O(N)/ O(KL) + O(logN) | O(2N+L$^2$)/ O(2NL+L$^2$) |

### E. Experimental Results

FWF Implementation Challenges

There are several challenges for the FWF implementation. The first issue is numeric instability and deals with the inverse of the correntropy matrix $V$ in (34). Large condition numbers will bias the solution and need to be corrected through regularization. The second issue stems from the fundamental fact that learning models must generalize well outside the training set. Note that there is no error in the FWF methodology, so this presents a different problem than in conventional machine learning where the regularization can be controlled by penalty terms in the cost function. In the FWF, generalization is controlled by the kernel size, and by the model order, the two hyper-parameters in the design. It is easy to see that small kernel sizes yield a correntropy matrix that approaches a scaled identity matrix, $aI$. This is because when kernel sizes are small, correntropy will peak when signals exactly match, and become very small when signals do not match. This increases specificity in the training set and also simplifies conditioning of the $V$ matrix, but it requires a large number of samples in the training set and a huge dynamic range in the computation to avoid losing information in the higher lags.

Therefore, small kernel sizes limit the number of lags that can be used in practice to represent the input space data correlations. Hence, in order to capture long time dependencies amongst the lags in a stationary signal, we must use larger kernel sizes in $\mathcal{H}_G$. However, if the kernel size is too large then the behavior of the correntropy function will approach the behavior of the auto-correlation function i.e., we lose the specificity provided by the higher order moments of the data PDF. The other drawback of employing larger number of lags is that the chances of ill-conditioning in the correntropy matrix increase. Hence, these trade-offs mean that kernel size selection and regularization of $V$ are vital for the performance of the FWF, and the kernel size becomes the key parameter for generalization.

Regularization of the Correntropy Matrix

We concluded that larger kernel sizes are needed to preserve information over the lags of the $V$ matrix. This means that $V$ will be more ill-conditioned, which can be quantified by the matrix's condition number. It is important to note that while regularization is helpful, we also need to control the number of lags to obtain optimal performance. The regularization of the $V$ matrix is depicted in equation (41). Our goal is to find a $\gamma$ such that the condition number of $V_{\text{reg}}$ is approximately equal to some desired condition number, which becomes a FWF hyper-parameter.

$$V_{reg} = V + \lambda I; \quad \lambda = \gamma . \min EigValue\ (V) \quad (41)$$

Using this framework, we found that condition numbers below 30 worked well, which is quite restrictive, but can be expected because we expect tiny errors in prediction to make FWF competitive with KAF approaches. These low condition numbers require a large amount of regularization, which unfortunately does not utilize all the information in the $V$ matrix affecting the accuracy of the FWF predictions.

Initial FWF Results: The Mackey-Glass Time Series

The Mackey-Glass (MG) times series is a chaotic time series, generated by

$$\frac{dx(t)}{dt} = -bx(t) + \frac{ax(t-\tau)}{1+x(t-\tau)^{10}}$$

The MG times series used in the following experiments was generated with $b = 0.1$, $a = 0.2$, and $\tau = 30$. Experiments testing the KLMS and KRLS kernel adaptive filters with this time series can be found in [14].

One of the hyper-parameters of the FWF is number of lags ($L$). This defines the length of the correlation time used to represent each sample, very similar to the Wiener model. Each sample is represented by a vector of length $L$ with the form $[x(i), \ldots x(i - L - 1)]^T$. This is standard practice for time series prediction. The second hyper-parameter is the kernel size of $\mathcal{H}_G$. To estimate the dependence of performance on hyperparameters, the parameters are scanned and plotted with training set data to obtain the performance surface of the FWF$_{LM}$, with two different local model orders (K = 5 and 15). We can see in Figure 1 that the two local model orders provide basically the same results. The minimum is obtained around L = 7 delays, and the minimum trough is around σ = 1.5, which is much larger than the corresponding KAF filters for the same time series. We also see that the best error is on the order of $10^{-3}$ (log 10) which is better than the Wiener filter of the same order for this data set (MSE = 0.013).

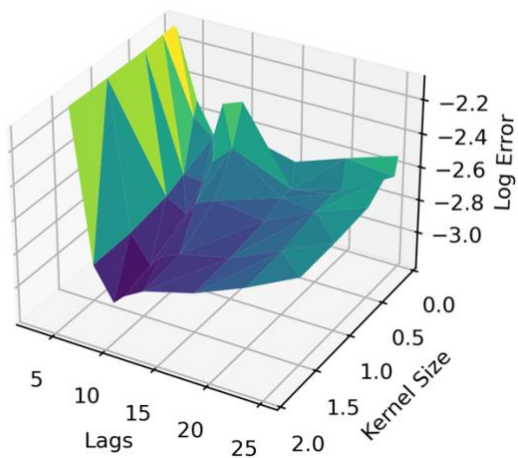 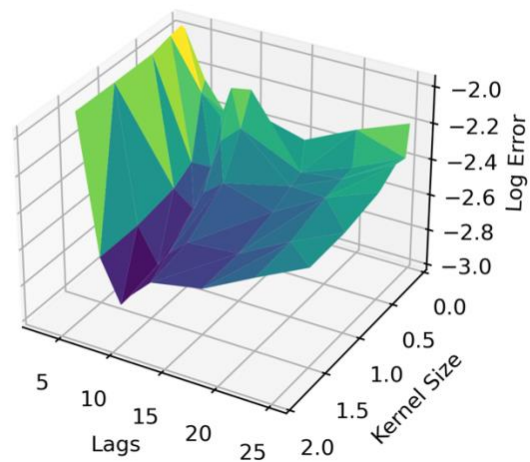

Figure 1. Error performance surface over the two FWF hyper-parameters (kernel size and number of lags), estimated with two different local model orders.

Experiments with FWF$_{FP}$ and FWF$_{LM}$

In this section, the performance of the FWF with both pre-imaging methods described above is compared with two well-known KAF methods, kernel recursive least squares (KRLS) and KLMS. Figure 2 compares the average test set MSE across 5-folds of cross validation. The best kernel size from Figure 1 was employed (σ =1.5). The figure shows performance with two different values of $K$ for the FWF$_{LM}$. We also present results with K=1 for a direct comparison with the FWF$_{FP}$. The number of lags considered for FWF$_{LM}$, was $L = 7$ the same as embedding for KLMS, and KRLS. The performance for the FWF$_{FP}$ is the worst, and it improves slightly with the number of lags, therefore the figures below show results with $L$=25. Notice that FWF$_{LM}$ with K=1 is much better than the fixed-point update and here rivals the performance for higher number of local models. Notice that, as expected, there is no variation with the number of local models in the FWF$_{FP}$ because the method uses an optimization to find the minimum, so the solution only depends on $L$, σ, and the number of samples in the training set. The FWF$_{LM}$ approaches the performance of KLMS, but it is far worse than KRL. Remember that the FWF was derived under a strict stationarity assumption, which is not fulfilled by the MG time series. Therefore, this result is quite reasonable, taking in consideration the FWF much smaller computation complexity.

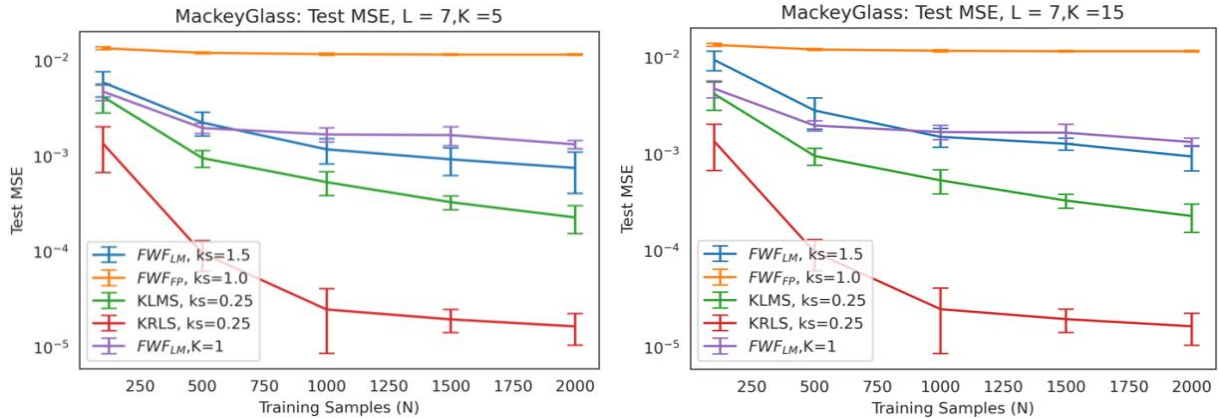

Figure 2. Comparisons of predictions for two different selections of local models (K) as a function of the number of samples in the training set. Asymptotic performance occurs after 1000 samples. For K=1 performance is much better than fixed point pre imaging. More models worsen the prediction results on MG.

**Noisy Mackey-Glass Prediction:**
In this experiment the FWF with both pre-imaging methods, KLMS and KRLS are predicting the MG time series, but with white Gaussian noise added to the input signal. Each algorithm is given a noisy training and testing input, and the desired signal is the next time point $x(t + 1)$ with no added noise. White Gaussian noise with standard deviations of 0.01, 0.04, 0.1, and 0.2 were tested. Five-fold cross validation was used for each algorithm at each noise level. The best kernel size is shown for each algorithm. In general, FWF$_{LM}$ is better than KLMS and KRLS at higher noise levels. The number of training samples does not have a great effect on the final MSE. Again, the performance of FWF$_{FP}$ is evaluated at $L = 25$ while FWF$_{LM}$ use $L = 5$ and 7.

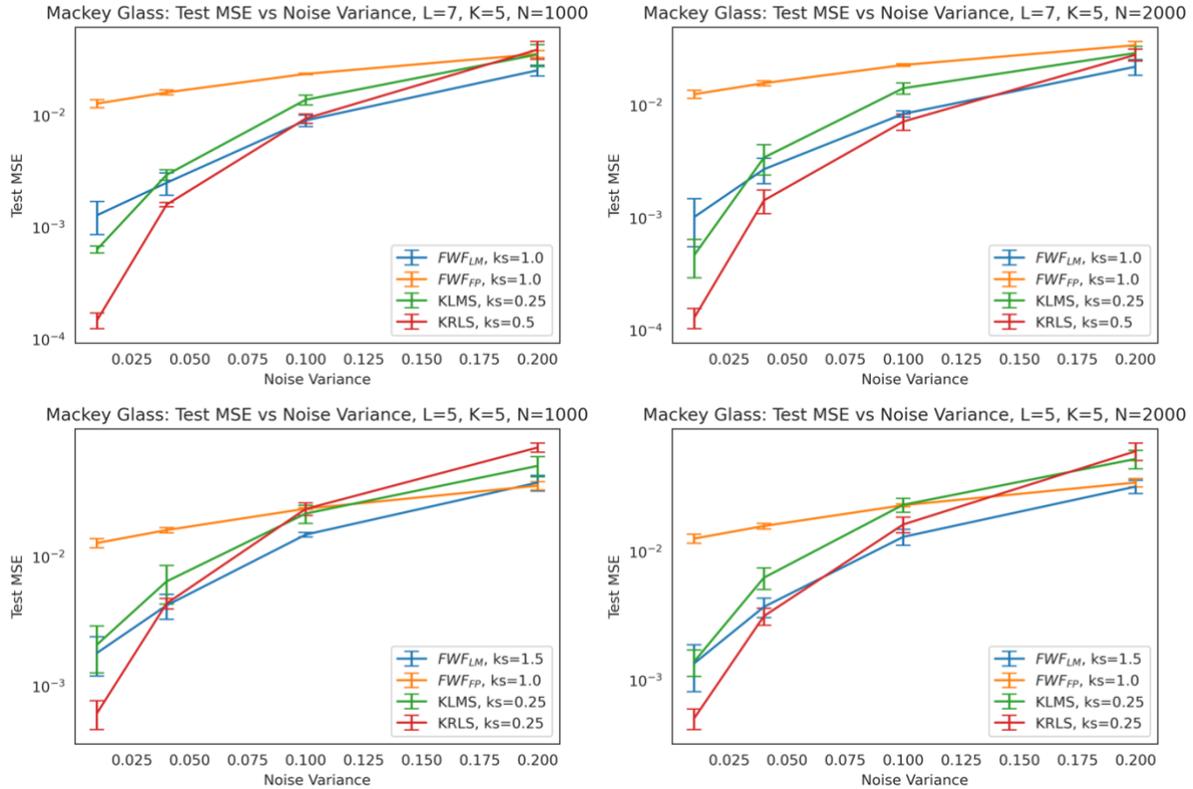

Figure 3. FWF has better robustness when noise is added to the time series, as we can expect from the use of multiple delays.

**Lorenz Prediction:**

We decided to test the performance of the FWF in the prediction of a more complex chaotic dynamical system. The Lorenz system is a well-known system introduced in [32]. We use the $x$ component of the Lorenz attractor and to make the problem harder, the model predicts $x(t + 10)$ e.g. 10 samples ahead with the last $L$ samples. A version of this experiment can be found in [13]. Like the previous experiments, the FWF$_{FP}$ was evaluated at $L = 30$, which is larger than the other methods. The FWF$_{LM}$ outperforms KLMS for low number of lags. This difference shrinks as we consider more lags. As in the other experiments, FWF$_{FP}$ does not perform well when compared to the other methods. In the Lorenz system, FWF$_{LM}$ performs at the level or better than the KLMS. Notice that this time series is far from stationary.

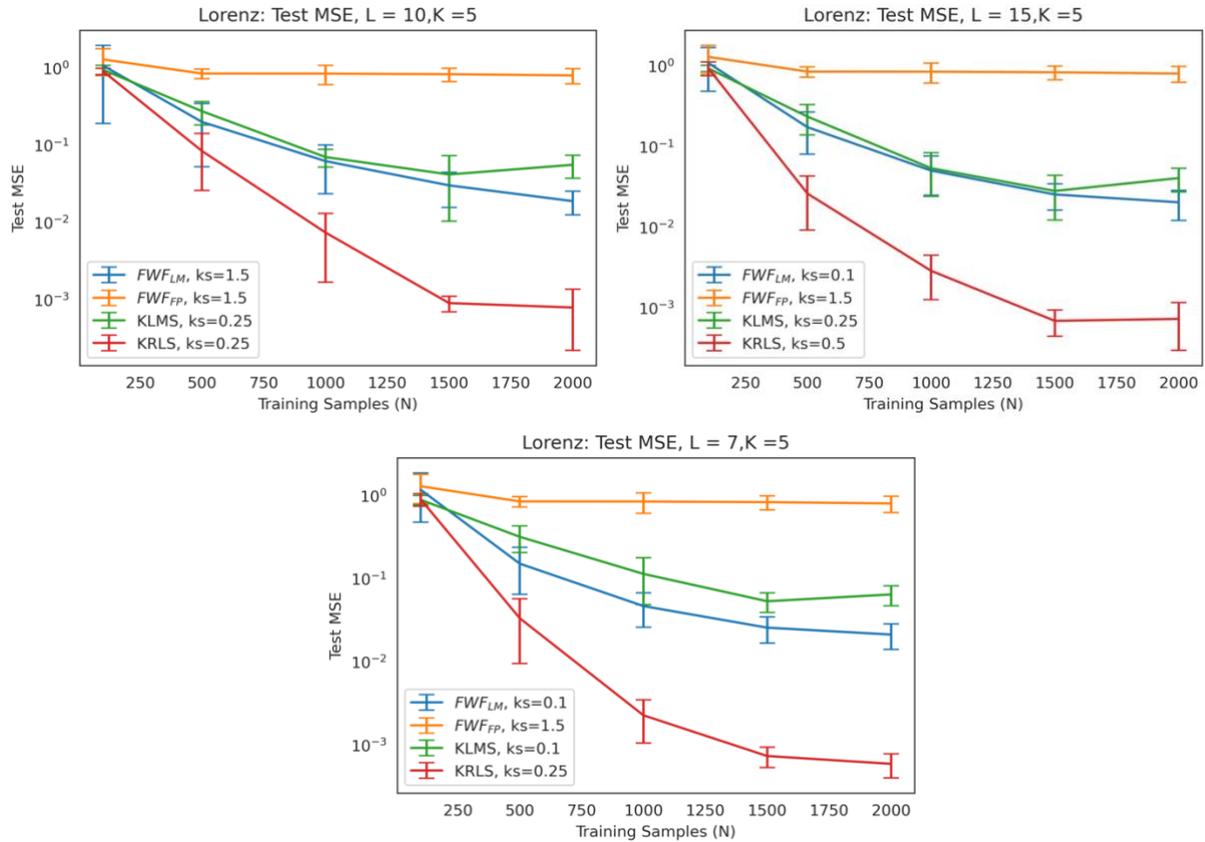

Figure 4. Comparison of performance in the Lorenz time series prediction. For this time series FWF$_{LM}$ performs better than KLMS but by a small margin.

**Further Analysis on Mackey-Glass sample by sample predictions**

Figures 5 shows the training and testing predictions compared to the desired with $L = 7$, kernel size of 1.5, and two different local models $K = 5$ and 100. In both, the prediction is worse in the parts of the Mackey-Glass series that are more non-stationary (the small ripple across the signal), but the smoothing effect of using many local models is clearly visible. This explains why K=1 does such a good job in this signal. This makes sense since when the model is more localized, the dependency on the stationarity constraint is reduced.

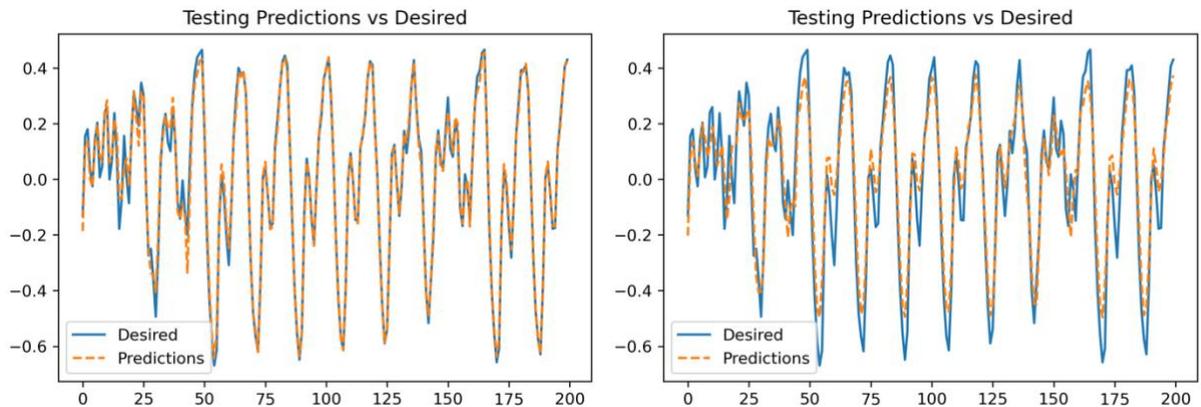

Figure 5. Sample by sample comparisons of predictions with the FWF$_{LM}$. Most of the errors occur in the time varying ripple superimposed in the signal. Notice that less local models perform better.

**Further Prediction Analysis on Lorenz:**

Figures 6 shows predictions made by the FWF$_{LM}$ on the Lorenz time series described in the above section. The hyperparameters here are $L = 7$, $\sigma = 0.1$, with two local models, of order $K = 5$ and $K = 100$. It is obvious that when the number of local models increases, samples too far away from the optimal solution will average out the response of the FWF, degrading the prediction. It is also interesting that the errors at the bottom of the signal ae not smooth, showing that there are not enough good neighbors in the training set.

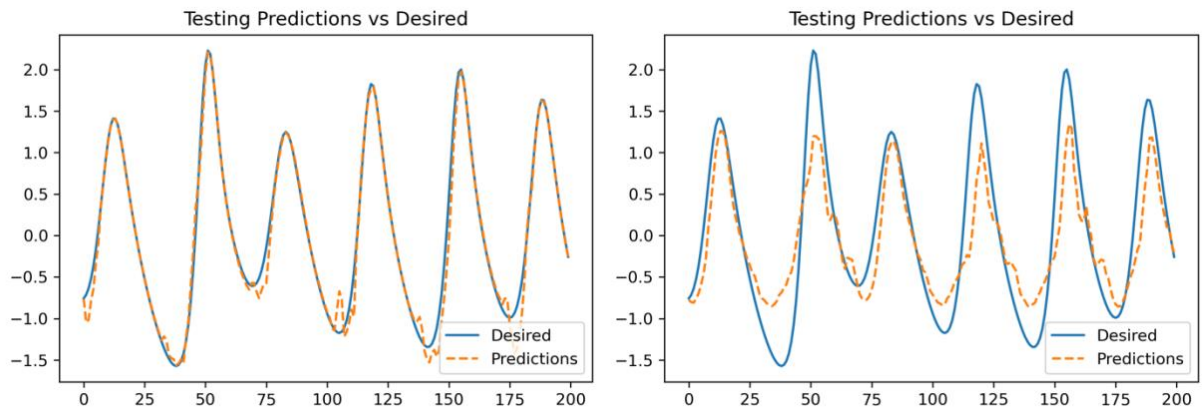

Figure 6. Averaging effect in the quality of the prediction when too many local models are employed (K=5 left, versus K=100 on the right).

### F. Conclusions

The main objective of this paper is to find a principled way to include the input data statistics in the inner product of a universal RKHS. Recall that KAFs use a data independent kernel (e.g. Gaussian) to project the data to define in the RKHS, the functional that implements the optimal model for the application. At test time for online applications, these functionals grow linearly with the number of samples, which is impractical. In practice, sparcification techniques must be used. The hypothesis is that a data dependent kernel will substitute the current KAF methodologies and simplify a lot the functional form to achieve an equally performing model. Parzen inspired this extension by showing that the ACF of a stationary random process is a positive definite kernel where optimal statistics models can be implemented. Once in this RKHS, a simple orthogonal projection is sufficient to find the optimal solution, unlike the incremental solution of KAF. However, the ACF kernel spans the input data space, so the RKHS solution is still a linear model with complexity higher than the Wiener filter. With this observation, the goal of this paper can be stated as extending Parzen's work to universal models.

The paper shows how to accomplish this task by defining the positive definite correntropy kernel as the inner product in a novel RKHS $\mathcal{H}_v$. The advantage is that functionals in $\mathcal{H}_v$ represent universal mapping functionals (for infinite number of lags), extending Parzen's result. The dimension of $\mathcal{H}_v$ is controlled by the number of delays of the autocorrentropy function, so this space is vastly different from the RKHS created by the Gaussian function, with the promise of decreasing the computational complexity of the implementation at test time. The paper presents the analytical solution of the FWF in $\mathcal{H}_G$, but we were unable to find a way to use the kernel trick to obtain the input space filter. The difficulty is that $\mathcal{H}_G$ is not congruent with $L_2$. Two pre-imaging techniques are proposed to implement the FWF in the input space, which are both approximated solutions, but they differ in the method and in the computation. FWF$_{FP}$ uses a fixed-point iteration to find the best solution to evaluate the functional in $\mathcal{H}_G$, but since one single Gaussian is unable to model well a sum of Gaussian at different centers, more sophisticated optimization methods are needed for good performance. The training set is never used in this pre-imaging technique. The FWF$_{LM}$ on the other hand uses the training set data to find pairs of samples that approach the best solution in the training set. This requires a search across the training set to find the best sample pairs to match the target response, but the method avoids the difficulty of FWF$_{FP}$ fixed-point iteration by averaging local models obtained in the training set. The simplest of the FWF$_{LM}$ with $K = 1$ may be applicable for many nonlinear applications. The FWF$_{LM}$ was found experimentally more accurate than the linear Wiener filter and is on par with the KLMS performance, but it is still substantially worse than KRLS. As an advantage, the FWF filter is far more efficient computationally than KAF implementations and uses less memory. Th FWF$_{FP}$ is of the same complexity as the Wiener filter but requires a recursive optimization at each iteration, which is not very expensive computationally. The FWF$_{LM}$ requires a search at the training time to match pairs of samples to find the pre image, and at test time, a search to find the closest training sample to the current test input (which is O(log N) if the input training samples are ordered by amplitude).

Hence, we conclude that this paper does not solve all the issues and should be considered as a first attempt to develop a new class of universal mappers in RKHS that integrate the data statistics in the kernel. But the novelty of the technique brings fresh ideas to statistical signal processing that need also to be further investigated. For instance, the FWF never employs the error, which is critical in KAFs. Effectively, FWF only works with the minimum norm (orthogonal) projection in RKHS, so it is "model agnostic": the important step is to create a RKHS that includes the data statistics (in the form of its autocorrentropy function) in the inner product. Of course, this construction requires "parameters" that are the ACF values of the input data and the CCF with the target, and the number of lags, just like least squares. After this construction, the FWF just finds the best local projection in the optimal RKHS functional centered at the current test sample. Therefore, there is no model nor parameters as in conventional optimal filtering and neural networks, just memory of the training set. In a sense, this approach resembles how brains encode and react to the physical world; neurons across life encode the structure and similarities given by the laws of physics, and they react very quickly to implement their response to stimulus, which means that the response must be very easy to compute. The advantage and disadvantages of the new approach are not fully understood at this time. Finally, we should focus on ways to avoid the loss of congruence between the universal RKHS and the input space. The correntropy RKHS has the very nice property that embeds the statistics of the data in the inner product, but there may be other kernels that maintain congruence with the input space, exemplified by our work and others on embedding PDFs in RKHS [33]. Another interesting aspect is that the local linear models seem to go beyond the strict stationarity assumption that supports theoretically the method. More work is required to study further this aspect.

Acknowledgements: This work was partially supported by ONR grants N00014-21-1-2295 and N00014-21-1-2345

## Appendix: Properties of the AutoCorrentropy Function

The existence of $\mathcal{H}_G$ opens new possibilities to extend the work of Parzen on the covariance RKHS that is defined on the Hilbert space of the data. Recall that the autocorrelation function of a time series is a similarity measure quantified by the expected value of the product between two random variables $X(t_i), X(t_j)$ at two different time intervals $t_i, t_j$ given by their joint distribution. As such it only measures the first moment (the mean) of the joint PDF over time. The first question is how to modify the autocorrelation function, as a similarity measure in such a way that it captures all the statistical information contained in the joint distribution.

### Going Beyond the Autocorrelation Function for Similarity

The most general measure of similarity in the joint space of two r.v. $X, Y$ is the cross covariance operator [26], defined by the bilinear form

$$C_{s,t}(f, g) = E[f(X)g(Y)] - E[f(X)] \cdot E[g(Y)] \quad (A1)$$

The covariance operator has been estimated in RKHS $\mathcal{H}_G$ as the matrix $\Sigma_{x_s x_t}$ of size equal to number of samples such that

$$\langle f, \Sigma_{x_t x_s} g \rangle_{\mathcal{H}_G} = C_{s,t}(f, g) \quad (A2)$$

where $f$ and $g$ are functional in RKHS that map the samples from the r.v. $x_t$ and $x_s$. But this treatment might be overly complicated for a stationary random process. Firstly, the marginals have the same density; secondly, only a scalar similarity over marginals is needed, and the mean embedding operator (20) can be estimated in $\mathcal{H}_v$; and thirdly because time establishes an a priori order on the r.v. such that a single variable (the delay) can be employed, instead of pairwise samples. Therefore, we submit that it is not necessary to estimate the full covariance operator for this application, which is computationally very intensive.

### Measures of Similarity in the Joint Space of Densities

*Definition*: Given a strictly stationary time series $\{X_t, t \in T\}$ the equality in probability density between two marginals at $s$ and $t$ i.e., $P(|X(s) - X(t)| < \varepsilon)$ for an infinitesimally small $\varepsilon$, defines a measure of similarity that can be estimated in $\mathcal{H}_G$.

In the joint space of $p_{s,t}(x_t, x_s)$ we can define a radial marginal as the bisector of the joint space. The density over the line $x_t = x_s$ approximates $\frac{P(|X(s) - X(t)| < \varepsilon)}{\varepsilon}$, which can be estimated as

$$E_{p_{s,t}}[\delta(X(s) - X(t))] \quad (A3)$$

where $\delta(.)$ is a delta function and we assume that the joint pdf over the lags is smooth along the bisector of the joint space is non-zero. To simplify, the Dirac calculus is used to illustrate the concept.

The expected value in (A3) can be written

$$E_{p_{s,t}}[\delta(X(s) - X(t))] = \iint \delta(x_s - x_t) p_{s,t}(x_s, x_t) dx_s dx_t \quad (A4)$$

The meaning of (A3) is quite clear: it is integrating the area under the joint density along the line $x_t = x_s$. Therefore, we can write (A4) as a single integral

$$E_{p_{s,t}}[\delta(X(s) - X(t))] = \int p_{s,t}(x, x) dx \quad (A5)$$

This reduction to a single integral can be expected by the definition of conditional PDF (see below), and it simplifies the calculation because of the statistical embedding in $\mathcal{H}_v$.

Note however, that this procedure needs to be repeated for every lag $L$ of interest i.e., it should be written as $t = s - l$, $l = 0, \ldots L$. Fortunately, the maximum lag $L$ is dictated by the embedding dimension of the real system that produced the time series, which is far smaller than the number of samples we collect from the world. In engineering applications this order can be estimated by Takens' embedding theory [27], or more practically by selecting the first minimum of the time series autocorrelation function. This computation is much simpler than the covariance matrix in (A1) because we are reducing the matrix to a vector $u$ of size $L$.

Correntropy functional as an approximation to the bisector integral

An empirical estimator of the natural measure of similarity defined above is given by its inner product (20). It turns out it has been coined in [16] the correntropy functional, which reads

$$V_\sigma(t, s) = E_{p_{t,s}}[G_\sigma(x_t - x_s)] \quad (A6)$$

where $G(.)$ is the Gaussian function with bandwidth $\sigma$. As discussed above, correntropy is a mean embedding of the joint pdf of a pair of samples. Rewriting (A6) using the definition of the expected value over the joint distribution, we obtain

$$V_\sigma(t, s) = \iint G_\sigma(x_t - x_s) p_{t,s}(x_t, x_s) dx_t dx_s = E[G_\sigma(x_t - x_s)] \quad (A7)$$

for strictly stationary processes. The best way to interpret this relation is to realize that when $x_t = x_s$, i.e. along the bisector of the joint space, the Gaussian kernel function is maximum, i.e. correntropy weights the joint space of samples with Gaussian kernels placed along the bisector of the first quadrant [16]. When the kernel size $\sigma$ approaches 0, it approximates a delta function $\delta(x_t - x_s)$, so we obtain an approximation to (A3). Moreover, correntropy is easily computed from samples too. Collect a segment of data of size $N$ from a time series. From (A7) an estimator of correntropy is simply

$$V_\sigma(\tau) = \frac{1}{N - \tau + 1} \sum_{i=m}^{N} G_\sigma(x_i - x_{i-\tau}) \quad (A8)$$

Hence, correntropy effectively estimates a radial marginal density obtained by integrating along the bisector from samples with linear complexity. This is unsuspected, because we are quantifying similarity in the structure of a time series beyond what we can achieve with the mean value of the product of samples in the autocorrelation. Note that here the kernel size should be made small for fine temporal resolution, but there is a compromise, because if we use a very small kernel size, the number of samples $N$ must be sufficiently large to get sufficient number of samples around the bisector of the joint space for accurate statistical estimation.

### The Relation between $P(x_{t_1} - x_{t_2})$ and the Conditional Density in the Joint Space

The Dirac calculus is a short cut and here we provide a more precise derivation of the value of the radial margin as a conditional distribution. As is well known the definition of conditional distribution of the r.v. $X$ given $Y$ is

$$f(x|y) = \frac{f(x,y)}{f(y)} = f(x|Y = y_0) = \frac{f(x, Y = y_0)}{f(Y = y_0)}$$

The meaning of this conditional is that we pick a value for $y = y_0$ and compute the area under the joint pdf at $y_0$. Here we are interested in a radial marginal, which is the bisector of the joint space given by the equality in probability i.e., $Y=X$, and would like to see how to compute it. Let us start with the distribution function and write the conditional probability as

$$F(x|(x - \delta) < Y \leq x) = P(X \leq x|(x - \delta) < Y \leq x) =$$
$$\frac{P(X \leq x, (x - \delta) < Y \leq x)}{P((x - \delta) < Y \leq x)} = \lim_{\delta \to 0} \frac{\int_{x-\delta}^{x} \int_{-\infty}^{x} f_{X,Y}(u,v) du dv}{\int_{x-\delta}^{x} f_Y(v) dv} = \frac{f_{X,Y}(x,x)}{f_Y(x)}$$

So, when the concept of the radial margin is employed as a conditional probability, we see that there is a normalizing factor that guarantees that the result adds to one as required for probabilities, but the numerator is exactly what the Dirac calculus quantifies in (A4).

### Approximating $P(x_{t_1} - x_{t_2})$ with Correntropy

$$\lim_{\sigma \to 0} v_\sigma(t_1, t_2) = \iint \delta(x_{t_1} - x_{t_2}) p_{p_{t_1} p_{t_2}}(x_{t_1}, x_{t_2}) dx_{t_1} dx_{t_2} = \int p_{p_{t_1} p_{t_2}}(x_{t_1}, x_{t_1}) dx_{t_1} \quad (A9)$$

In practice, the kernel size is always finite so correntropy does not estimate the probability density over a line in the joint space but the probability on a "Gaussian shaped tunnel" of width $\sigma$ along the radial direction $x_{t_1} = x_{t_2}$, which will be approximated by a parallelepiped of width $2\varepsilon$ with $\varepsilon \sim 1.25\sigma$. We can write

$$P(|x_{t_1} - x_{t_2}| < \varepsilon) = \int_{x_{t_1}=-\infty}^{\infty} \int_{x_{t_2}=x_{t_1}-\varepsilon}^{x_{t_1}+\varepsilon} p_{p_{t_1} p_{t_2}}(x_{t_1}, x_{t_2}) dx_{t_1} dx_{t_2} \quad (A10)$$

If $\varepsilon$ is small and $p_{p_{t_1} p_{t_2}}(x_{t_1}, x_{t_2})$ is continuous at every point along the $x_{t_1} = x_{t_2}$ line, the function value does not change a lot along $x(t_2)$ within the interval $[x_{t_1} - \varepsilon, x_{t_1} + \varepsilon]$ for any fixed $x(t_1)$. Thus

$$P(|x_{t_1} - x_{t_2}| < \varepsilon) \approx 2\varepsilon \int_{x(t_1)=-\infty}^{\infty} p_{p_{t_1} p_{t_2}}(x_{t_1}, x_{t_1}) dx_{t_1} = 2\varepsilon v_\sigma(t_1, t_2) \quad (A11)$$

And finally, we have

$$v_\sigma(t_1, t_2) = \frac{P(|x_{t_1} - x_{t_2}| < \varepsilon)}{2\varepsilon} \quad (A12)$$

which shows that correntropy estimates indeed the *probability density* of the event $P(x_{t_1} = x_{t_2})$ in the joint sample space for small kernel sizes.